Title:

# Unsteady self-sustained detonation waves in flake aluminum dust/air mixtures

Authors:

Qingming Liu[1], Jinxiang Huang[1], Yunming Zhang[1,2], Shuzhuan Li[1]

Affiliation:

[1]State Key Laboratory of Explosion Science and Technology

Beijing Institute of Technology, Beijing, 100081, China

[2]Department of Fire Protection Engineering

Chinese People's Armed Police Force Academy

Corresponding author: Qingming Liu

Address: No. 5 South Zhongguancun Street, Haidian District, Beijing 100081, China

Telephone number: 86-10-68914261

Fax number: 86-10-68914287

E-mail address: qmliu@bit.edu.cn;

# Unsteady self-sustained detonation waves in flake aluminum dust/air mixtures

Abstract

Self-sustained detonation waves in flake aluminum dust/air mixtures have been studied in a tube of diameter 199 mm and length 32.4 m. A pressure sensor array of 32 sensors mounted around certain circumferences of the tube was used to measure the shape of the detonation front in the circumferential direction and pressure histories of the detonation wave. A two-head spin detonation wave front was observed for the aluminum dust/air mixtures, and the cellular structure resulting from the spinning movement of the triple point was analyzed. The variations in velocity and overpressure of the detonation wave with propagation distance in a cell were studied. The interactions of waves in triple-point configurations were analyzed and the flow-field parameters were calculated. Three types of triple-point configuration exist in the wave front of the detonation wave of an aluminum dust/air mixture. Both strong and weak transverse waves exist in the unstable self-sustained detonation wave.

## 1 Introduction

Flake aluminum is an important raw material in industry. It has been widely used as an additive in paints, fuels, propellants, and other energetic materials. Aluminum dust explosion is one of the most serious dust explosion hazards due to the high heat release of the combustion reaction. In some circumference, such a dust explosion can develop into detonation and cause more serious damage to human life and property. The minimum ignition energy, lower and upper concentration limits, and the violence of dust explosion have been studied extensively [1, 2].

Detonation waves in oxygen/carbon monoxide and oxygen/acetylene mixtures in a round tube of diameter 14 mm have been studied by Voitsekhovskii, Mitrofanov, and Topchian [3], who employed a compensating photography technique to record the luminosity of the flame front. Soot foil was used to record the trajectory of the triple point and piezoceramic gauges were used to capture the pressure profiles of the detonation wave front. Spinning detonation waves were identified and the wave structure was analyzed. Spinning detonation waves in acetylene/oxygen mixtures highly diluted with argon have been studied in circular tubes of diameters 1 inch (2.5 cm) and 4 inch (10 cm) by Schott [4]. Combining the results of simple photographic, electrical, and mechanical measurements, the wave structures of the detonation wave were deduced. The cellular structures of the detonation waves in acetylene/oxygen/argon mixtures have been studied by smoke-foil and emission techniques in a 25.5 mm inner diameter round tube [5], and the cell lengths were determined by analyzing the soot imprint and the period of the emission signals. By using a planar laser fluorescence technique, the distribution of OH radicals resulting from the detonation reaction in hydrogen/oxygen/argon mixtures was measured by Pintgen et al. [6, 7]. The keystone structure result from the triple-shock Mach interactions was observed. Diverging and converging cylindrical detonation waves in gaseous fuel/air mixtures, with wave fronts appearing as cylindrical surfaces, have been studied by Lee and Knystautas [8, 9], and it was found that the trajectories of the transverse waves showed two intersecting sets of logarithmic spirals. Spherical detonations in gaseous fuel/air mixtures have been studied by Shchelkin and Troshin [10], Duff and Finger [11], and Lee et al. [12]. Soot prints of the spherical detonations showed identical cellular structures to planar detonation waves propagating in a tube for the corresponding mixture. Cell size data of detonation waves for most common gaseous fuel/air mixtures or fuel/oxygen/inert diluent mixtures under various initial conditions have been collected and are available at the Caltech Explosion Dynamics Laboratory website of J.E. Shepherd [13]. Studies of detonation in dust/air mixtures have been conducted in experimental tubes of diameters 141 and 122 mm, respectively [14, 15]. Two-head spinning detonation was found in corn starch/oxygen mixtures [15]. The deflagration-to-detonation transition (DDT) in flake aluminum dust/air mixtures has been studied by Liu, Li, and Bai in a 199 mm inner diameter experimental tube [16]. The DDT distance was found to be approximately 51 times the tube diameter. A self-sustained detonation wave was formed and a single-head detonation wave with a pitch size of 495 mm was observed.

In the present work, the characteristics of self-sustained detonation waves in flake aluminum dust/air mixtures have been studied and analyzed. A specially constructed experimental test section with 118 sensor mounting holes in the tube wall has been used to study the wave structure of the self-sustained waves. A two-head spinning detonation wave front was observed and the characteristics of the wave front have been analyzed. The interactions among the primary shock wave, transverse wave, and Mach reflection wave have been studied.

2. Experimental

2.1 Experimental set-up

The experimental system is a Horizontal Multiphase Combustion and Explosion System (HMCES) in the State Key Laboratory of Explosion Science and Technology, Beijing Institute of Technology, and is shown in Fig. 1. The inner diameter (ID) of the horizontal multiphase combustion and explosion tube is 0.199 m, the total length is 32.8 m, with a multiphase experimental section of length 30.8 m and a transition section of length 1.6 m. The effective L/D ratio is 155. The system mainly comprises a multiphase experimental section, incorporating a photo access test section and a detonation wave structure test section, a transition section, a dumping tank, a set of 44 dispersion units, a control unit, an ignition system, a pressure measuring system, and other equipment. One end of the multiphase experimental section is sealed with a flange, while the other end is connected to the transition section, which is in turn connected to a dumping tank of volume 13 $m^3$. The 44 dispersion units are mounted on either side of the multiphase experimental section at regular intervals of 0.7 m. The length of the photo access testing section, with two high-strength optical glass windows on opposite sides of the tube wall, is 1.4 m. The length of the detonation wave structure test section, with 118 sensor mounting ports arranged on 15 circumferences of the tube wall, is 1.4 m. A plastic film is placed between the multiphase experimental section and the transition section to form a uniformly dispersed dust cloud. The detection holes are arranged along the wall of the tube at a spacing of 0.7 m. A total of 17 Kistler pressure sensors are located on the inner wall surface of the horizontal combustion and explosion tube. The formation and development of the compressive wave induced by dust combustion and explosion, the formation and strengthening of the shock wave, and the transition from deflagration to detonation in dust/air mixtures can be detected by the pressure sensors arranged along the axial direction of the experimental tube. The distances of the sensors from the closed end of the tube are 1.75, 3.15, 4.45, 6.65, 8.05, 9.45, 10.85, 12.25, 13.65, 15.05, 17.15, 19.25, 21.35, 23.45, 25.55, 26.65, and 29.61 m, respectively. Moreover, 32 Kistler pressure sensors arranged around four circumferences at a separation of 0.14 m are deployed in the detonation wave structure test section, through which self-sustained detonation passes in the present study. About each circumference, there are 8 sensors at 45° intervals. The first circumference with pressure sensors is located at 29.96 m from the closed end of the tube. This arrangement of the pressure sensors allows one to study the deflagration-to-detonation transition (DDT) and to deduce the wave structure of the self-sustained detonation wave propagated in multi-phase combustion mixtures. The pressure measurement system consists of Kistler piezoelectric sensors, an adapter, and a data acquisition system. The dispersion system consists of hemispherical nozzles, a sample chamber, a direction valve, a solenoid valve, a high-pressure air chamber, and a manual valve. The nozzles are connected to the sample chamber, which is in turn connected to the high-pressure air chamber and the manual valve via the solenoid valve. Finally, the manual valve is connected to an air compressor. Dispersion, ignition, and triggering of the data acquisition system are all controlled by the control unit. An electric spark of energy 40 J is used to ignite the aluminum dust/air mixtures in the experimental tube. The whole experimental system has been documented previously [16], and its reliability has been established.

The pressure used to disperse the aluminum dust is 0.8 MPa. The time interval between the start of dispersion of the aluminum dust and the start of ignition of the aluminum/air mixtures is 370 ms. The ambient pressure of the experiments performed in the tube is 0.14 MPa.

The aluminum dust used for the experiments had a flaked appearance, and a coating surface of 0.7 m$^2$/g. The activated aluminum content was no lower than 82%, and the particle size was distributed between 15 and 40 μm.

2.2 Initiation of detonation

The detonation wave in an aluminum dust/air mixture was initiated by the DDT process of the same aluminum/dust air mixture in the experimental tube. Using the experimental tube described above, the DDT and detonation wave structure in flake aluminum dust/air mixtures have been studied. By injecting aluminum dust into the experimental tube, aluminum dust/air mixture clouds formed inside the tube. Controlled by the control unit, the cloud was ignited by an electric spark ignition device, and the flame propagated in the experimental tube. Typical results of variations in peak pressure and velocity of shock/reaction wave with propagation distance in flake aluminum dust/air mixtures are shown in Fig. 2. With the acceleration of the flame in the tube, a compression wave formed and was amplified with its propagation along the tube. With the positive feedback between the compression wave and turbulent flow, the turbulent flame was accelerated, a shock wave formed, and this was strengthened with its propagation along the tube. With the acceleration of the shock wave, the flame and the shock coupled and a self-sustained detonation wave formed at 14 m. From Fig. 2, it can be seen that the DDT distance in the aluminum dust/air mixture was 14 m, and that the self-sustained detonation wave was characterized by oscillations in its propagating velocity and peak overpressure. The averaged propagating velocity of the self-sustained detonation in the flake aluminum dust/air mixture was 1570 m/s, with the peak overpressure ranging from 3.5 MPa to 7.0 MPa.

3. Results and discussion

3.1 Interactions of shock waves and traces of the triple points

Initiated by the DDT process, the onset of the detonation wave in an aluminum dust/air mixture with a dust concentration of 643 g/m$^3$ took place at 14 m from the ignition point. It can be seen from Fig. 2 that from the onset of the detonation wave at 14 m to 17.15 m, the velocity of the detonation wave was maintained at approximately 1570 m/s while the overpressure slightly decreased. From 17.15 m to the end of the experimental section, overpressure oscillation began, and the self-sustained detonation with a transverse wave moved along the tube wall in spiral manner. Since the distance between pressure sensors was greater than the cell size of the detonation wave, the peak values of overpressure and velocity of the wave front may not have been captured. From the pressure sensors on the top side of the tube in the wave structure test section, the velocity of the detonation oscillated between 1450 m/s and 1760 m/s, as shown in Fig. 9. The pressure histories on the periphery of the detonation wave front at circumferences 1–4 are shown in Figs. 3–6. Using the averaged velocity and the arrival time of the detonation wave at circumferences 1–4, the shapes of the shock fronts, with the angle on the periphery at each circumference, could be determined, as shown in Fig. 7, and a triple point (cusp) resulting from the interaction of the incident shock wave, the transverse wave, and Mach stem appeared in the curves of the shock-wave fronts. The variations in peak overpressure at the periphery of the detonation wave with the angle at each circumference are shown in Fig. 8. From Figs. 7 and 8, it can be seen that the peak overpressure appears at the triple point of the wave front at each circumference. The trajectory of the triple point shows a regular cellular structure. The peak overpressure and the direction of movement of the triple point at each circumference are shown in Fig. 8. The variations in the velocity and overpressure of shock waves with propagation distance in a cell are shown in Fig. 9. The velocity of the shock wave front in a cell is calculated from the axial spacing between the pressure sensors and the time interval required for the shock wave to pass through it. Hence, the maximum velocity may not be captured because of the large spacing between pressure sensors. The results indicated that at the beginning of the cell, immediately after the collision of the two transverse waves, an explosion center was produced, and the averaged shock-front velocity was greater than 1760 m/s, with a corresponding peak overpressure of 10.0 MPa. With the propagation of the leading shock front, its velocity decreased while the peak overpressure increased. The peak overpressure of the wave front attained its maximum value of 12.5 MPa and the averaged velocity of the shock wave decreased to 1545 m/s at a distance of one-quarter of the cell length. Subsequently, the velocity and overpressure decreased to lower values. At the end of the cell length, the peak overpressure and shock velocity of the wave front had decreased to 6.5 MPa and 1450 m/s, respectively. From Fig. 7, the track angle of the two-head spin detonation wave can be determined. The triple point moved 280 mm in the axial direction, while it moved by half of the perimeter of the tube in the circumferential direction. Hence, the track angle of the spinning detonation was 48°. The velocity of the transverse wave in the circumferential direction was 1744 m/s and cell length of the two-head spin detonation wave was 560 mm. The movements and periodical collisions of the transverse waves in the detonation wave produced an explosion center in the fuel/air mixture and thus changed the detonation parameters (velocity, overpressure, and temperature) and the detonation cell size, as shown in Fig. 9.

The Mach stem and the primary shock wave can be recognized by using the shapes of the shock front, the movement of the triple points, and the variations of shock velocity and overpressure with propagation distance in a cell, as shown in Fig. 7. The letter M denotes the Mach stem, while the letter I denotes the incident shock wave.

3.2 Configurations and flow-field parameter calculations of the triple point in the detonation wave in flake aluminum dust/air mixtures

Using the values of the track angle α and the shapes of the shock front at the periphery of the detonation wave, the flow angle of the primary shock wave $\varnothing$ can be determined. All of the parameters of the flow field around the leading front triple point A can be calculated by using the track angle α, the average detonation velocity D, and the flow angle. Following the classical oblique shock theory, all parameters behind the shock (pressure, temperature, density, and velocity) can be computed from the shock velocity, the incident flow angle $\emptyset$, and the initial conditions ahead of the shock. Typical triple-point configurations are shown in Figs. 11 and 12.

The dynamic equations for the calculation are:

$$\frac{P_b}{P_a} = \frac{2\gamma}{\gamma+1}\left(M^2\sin^2\emptyset - \frac{\gamma-1}{2\gamma}\right) \quad (1)$$

$$\frac{\rho_b}{\rho_a} = \frac{(\gamma+1)M^2\sin^2\emptyset}{(\gamma-1)M^2\sin^2\emptyset+2} \quad (2)$$

$$\frac{u_b}{u_a} = \frac{(\gamma-1)M^2\sin^2\emptyset+2}{(\gamma+1)M^2\sin^2\emptyset}\frac{\sin\emptyset}{\sin\varphi} \quad (3)$$

$$\frac{T_b}{T_a} = \frac{2\gamma}{\gamma+1}\left(M^2\sin^2\emptyset - \frac{\gamma-1}{2\gamma}\right)\frac{(\gamma-1)M^2\sin^2\emptyset+2}{(\gamma+1)M^2\sin^2\emptyset} \quad (4)$$

$$\tan\varphi = \tan\emptyset\frac{(\gamma-1)M^2\sin^2\emptyset+2}{(\gamma+1)M^2\sin^2\emptyset} \quad (5)$$

where the subscripts a and b refer to the quantities ahead of and behind the shock front, respectively. M is the Mach number of the incident shock and $\varphi$ is the angle of the flow behind the shock. If the temperature behind the shock wave is high enough, the chemical reaction of the fuel/air mixture can be considered as practically instantaneous. The parameters behind a shock wave with instantaneous reaction can be calculated according to the parameter relationships for an oblique detonation wave as follows:

$$P_b = \frac{\rho_a(u_a \times \sin\emptyset_3)^2}{\gamma+1} \quad (6)$$

$$\frac{\rho_b}{\rho_a} = \frac{\gamma+1}{\gamma} \quad (7)$$

$$\tan\varphi = \frac{\gamma}{\gamma+1}\tan\emptyset \quad (8)$$

$$u_b = \frac{\gamma u_a}{(\gamma+1)}\frac{\sin\emptyset_3}{\sin\varphi_3} \quad (9)$$

$$\frac{T_b}{T_a} = \frac{P_b}{P_a}\frac{\rho_b}{\rho_a} = \frac{\rho_a(u_a \times \sin\emptyset_3)^2}{P_a}\frac{\gamma}{(\gamma+1)^2} \quad (10)$$

The calculation procedures to determine the triple-point configuration and flow-field parameters of the detonation wave in an aluminum dust/air mixture are shown in Fig. 10. As shown in Fig. 10, on the contact/slip surface in the triple-point configuration, the pressures on opposite sides of the contact surface should be balanced, and the velocities on either side of the slip surface should be parallel to one another. The differential angle between the inlet flow angle $\varnothing$ and the outlet flow angle $\varphi$ of the shock wave can be expressed as: $\theta = \varnothing - \varphi$. The flow-field velocities on either side of the slip surface AD should satisfy one of the following criteria:

$$\theta_2 = \theta_1 + \theta_3; \theta_1 = \theta_2 + \theta_3; \theta_3 = \theta_2 + \theta_1;$$

and the flow-field velocity on either side of the slip surface DF should satisfy one of the following relationships:

$$\theta_4 = \theta_2 + \theta_5; \theta_4 = \theta_2 - \theta_5; \theta_4 = \theta_5 - \theta_2;$$

Each criterion corresponds to a triple-point configuration. Three triple-point configurations may exist at the first and second triple points. Based on the shapes of the shock-wave front, the track angle of the triple point, and the velocity of the detonation wave, the configurations and flow-field parameters of the triple point in a self-sustained detonation wave in an aluminum dust/air mixture were calculated. Three kinds of double triple-point configurations were obtained in detonation waves in aluminum dust/air mixtures, as shown in Figs. 11 and 12. The flow fields of double triple points with inlet flow angles of 28° and 32.5° are shown in Figs. 11 and 12, respectively. The results of the calculation showed that the temperature behind AA1 and BC in the double triple-point configuration shown in Fig. 11 was so great that the chemical reaction can be considered as practically instantaneous. Therefore, these shocks were taken to be detonation waves in the computations. When the inlet flow angle of the shock is small, the temperature after the shock front is not high enough for the chemical reaction to take place instantaneously. Hence, the parameters behind the shock are calculated by using the parameter relationships of the shock wave. With an increase in the incident angle, the temperature behind the incident shock increased abruptly, the chemical reactions in the fuel/air mixture behind the incident shock took place instantaneously, and so the parameters behind the shock could be calculated by the parameter relationships of a detonation wave, as shown in Fig. 12. In the flow configurations discussed, the greatest pressure exists behind waves BD and BC. In the flow-field configuration S1 shown in Fig. 11 (a), the angle between shock waves AB and BC is about 82°. The highest pressure is 12.2 MPa, which is much greater than that calculated by CJ theory [16]. In this flow configuration, the fluids behind waves AA1 and BD are compressed due to their convergence. The wave system DFG is a shock wave with a slide surface. In the configuration S2 shown in Fig. 11 (b), the angle between shock waves AB and BC is 158°. The greatest overpressure is 7.2 MPa, which is much lower than that in configuration S1. This can be explained in terms of the divergent flow of the fluids behind waves AA1 and BD. In configuration S2, DFG is a centered rarefaction wave. The flow-field configurations and parameters at triple point A are same for configurations S1 and S2. The strength of the transverse wave AB at point A can be expressed by S [17]: $S = \frac{P_b}{P_a} - 1$.

According to calculation by the above equation, the strength of the transverse wave at point

A is 0.41. The differences in the configurations S1 and S2 result from the structure at the triple point B. The strengths of the transverse wave BD at triple point B in configurations S1 and S2 are 2.7 and 1.2, respectively, with maximum overpressures of 12.2 and 7.2 MPa, respectively. The transverse wave structure in configuration S1 may be regarded as a strong transverse structure, while that in configuration S2 may be regarded as a weak transverse structure [17]. With increasing inlet flow angle, the pressure and temperature increased rapidly. Fast chemical reaction took place in the shock wave front AA2, so the parameter relationships of shock AA2 should be replaced by detonation relationships described by Equations (6)–(10). The flow-field configuration of the double triple point with an inlet flow angle of 40° in a flake aluminum dust/air mixture is shown in Fig. 12. In triple-point configuration S3 shown in Fig. 12, the strength S of the shock wave AB is 1.46 and the pressure behind the shock wave AA1 is 4.1 MPa. At the second triple point B, the angle between BC and AB in configuration S3 is 130°. The strength S of shock wave BD in configuration S3 is 0.4. The wave system DFG in configuration S3 is a centered rarefaction wave. The peak overpressure in configuration S3 is 5.7 MPa. The structure of the transverse wave shown in configuration S3 can be classified as weak transverse wave structure.

The calculated results for the triple-point configurations and the parameters of the detonation wave in an aluminum dust/air mixture at an initial pressure of 0.14 MPa with a self-sustained detonation of 1570 m/s and a track angle of 48° are summarized in Table 1. The structure of the detonation wave in the aluminum dust/air mixture in a round tube is shown in Fig. 13. The solid line represents the shock front, the dashed line represents the slide surface, and the dash-dotted line represents the flame. Numbered lines represent gauge trajectories relative to the wave front. From Figs. 3–6 and 13, and the calculated results of the triple-point parameters, it was found that near the triple point, the shock and flame were closely coupled, overdriven detonation occurred, and the maximum overpressure was attained just behind the transverse waves BD and BC. On moving away from the triple point, the flame became detached from the shock, and the peak overpressure of the shock decreased. The abrupt changes in overpressure and temperature were caused by the triple-point wave structure.

4 Conclusions

Self-sustained detonation waves in flake aluminum dust/air mixtures initiated by the DDT process have been studied in a tube of diameter 199 mm and length 32.4 m. The self-sustained detonation wave was characterized by the oscillation of peak overpressure with propagation distance. A two-head spin detonation wave was found in an aluminum dust/air mixture with an aluminum concentration of 643 g/m$^3$, the average velocity of the self-sustained detonation wave was 1570 m/s, and the overpressure ranged from 2.8 MPa to 12.15 MPa. The cell size of the two-head sustained detonation in this aluminum dust air mixture was 560 mm and the track angle of the triple point was 48°.

The structure of the detonation wave front in the aluminum dust/air mixtures was studied by means of a pressure sensor array mounted around the circumference of the tube. A two-head spin detonation wave formed in the aluminum dust/air mixture and a cellular structure resulting from the movement of the triple point was obtained. At the beginning of the cell, resulting from the collision of the transverse wave, an overdriven detonation wave was formed. With the propagation of the detonation wave in a cell, its velocity decreased while its peak overpressure increased in a quarter of the length of the cell. After a quarter of the cell size, the velocity and overpressure of the detonation wave had decreased and the flame and the shock had been decoupled. The maximum transient velocity of the detonation wave at the beginning of the cell was no lower than 1750 m/s and the minimum velocity of the detonation at the end of the cell was about 1450 m/s.

The shock-wave interaction at the triple point of the detonation wave was analyzed and the flow-field parameters were calculated. Based on the shapes of the wave front and the velocity of the detonation wave, three types of triple-point configuration were obtained and the corresponding flow-field parameters were calculated. The calculated pressure results are in good agreement with the experimental values. By analyzing the strengths of the shock waves and the shapes of the triple-point structures, strong and weak transverse waves were identified in the flake aluminum dust/air mixtures.

Figure and Table captions

Fig. 1. Schematic diagram of a horizontal multiphase combustion and explosion system.

1. experimental tube, 2. dispersion system, 3. ignition device, 4. control unit, 5. pressure sensor, 6. data acquisition system, 7. photo access test section, 8. shock wave test section, 9. air compressor, 10. plastic film, 11. venting system, 12. transition section, 13. dumping tank.

Fig. 2. Variations of peak pressure and velocity with propagation distance.

Fig. 3. Pressure histories at the tube circumference 1 in flake aluminum dust/air mixture.

Fig. 4. Pressure histories at the tube circumference 2 in flake aluminum dust/air mixture.

Fig. 5. Pressure histories at the tube circumference 3 in flake aluminum dust/air mixture.

Fig. 6. Pressure histories at the tube circumference 4 in flake aluminum dust/air mixture.

Fig. 7. Shapes of shock wave front with angle at circumferences 1−4 and traces of triple points on tube wall in aluminum dust/air mixture.

M − Mach reflection wave, I − Incident wave, T1−T4 labels of triple point

Fig. 8. Variations of peak overpressure with angle.

Fig. 9. Variations of propagation velocity and peak overpressure with propagation distance in a cell.

Fig. 10. Flow chart for the calculation of the configuration and parameters of triple-point structure.

Fig. 11. Strong (S1) and weak (S2) transverse wave structures and triple-point flow parameters of detonation waves in aluminum dust/air mixtures at an inlet flow angle of 32.5°.

S1 - strong transverse wave structure, S2 - weak transverse structure.

AA2 - incident shock wave, AA1 - Mach reflection wave, AB and BD - oblique shock waves, BC - transverse detonation wave, DFG - shock wave and slip surface (S1) or centered rarefaction wave (S2).

Fig. 12. Weak transverse wave structures and triple-point parameters of detonation waves in an aluminum dust/air mixture at an inlet flow angle of 40°. S3 - weak transverse structure.

Fig. 13. Wave-front structures of self-sustained detonation waves in aluminum dust/air mixtures.

Table 1.

Configurations and flow-field parameters of the two-head spinning detonation in aluminum dust/air mixture under initial pressure $P_0$=0.14 MPa, with a self-sustained detonation velocity of 1570 m/s and a tracking angle of 48°.

Figures

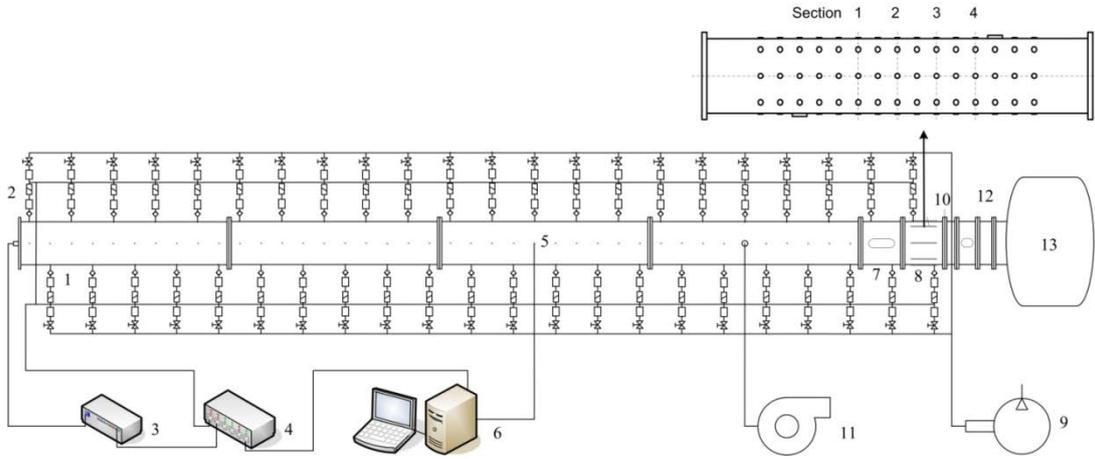

Fig. 1

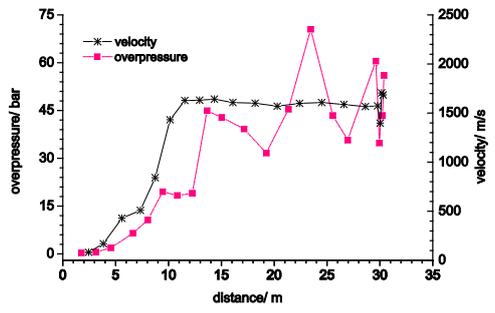

Fig. 2

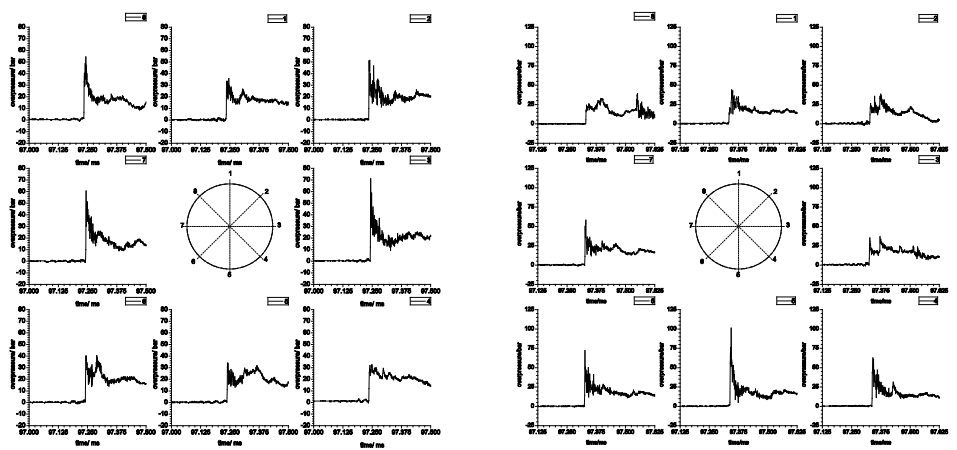

Fig. 3    Fig. 4

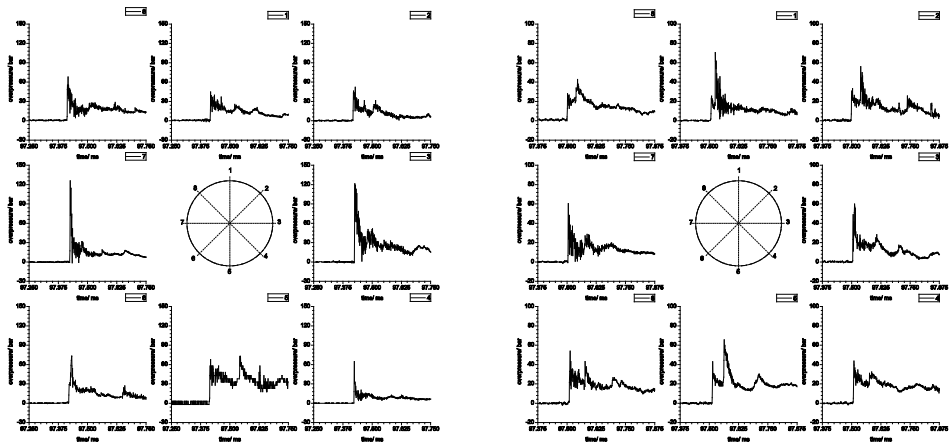

Fig. 5    Fig. 6

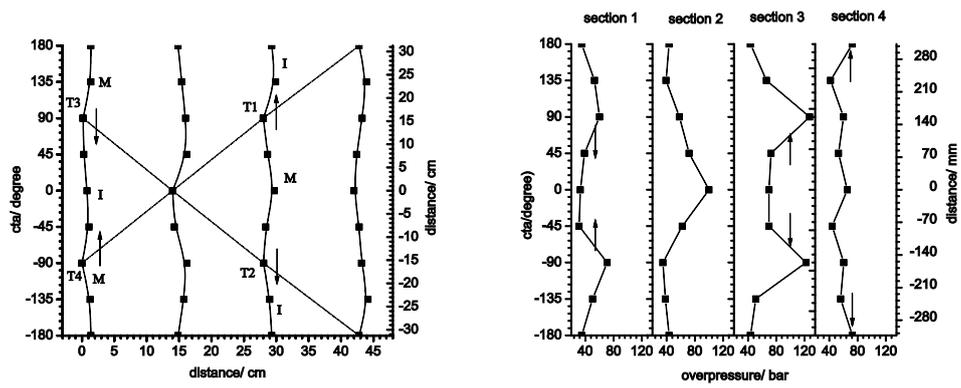

Fig. 7    Fig. 8

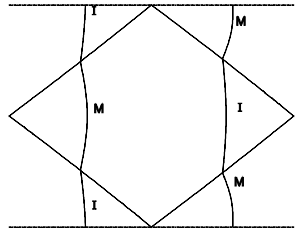

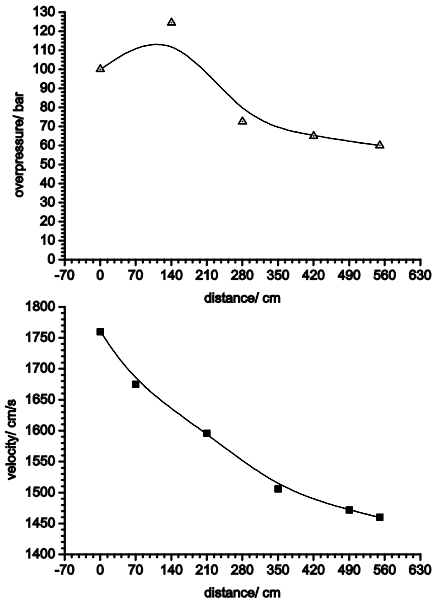

Fig. 9

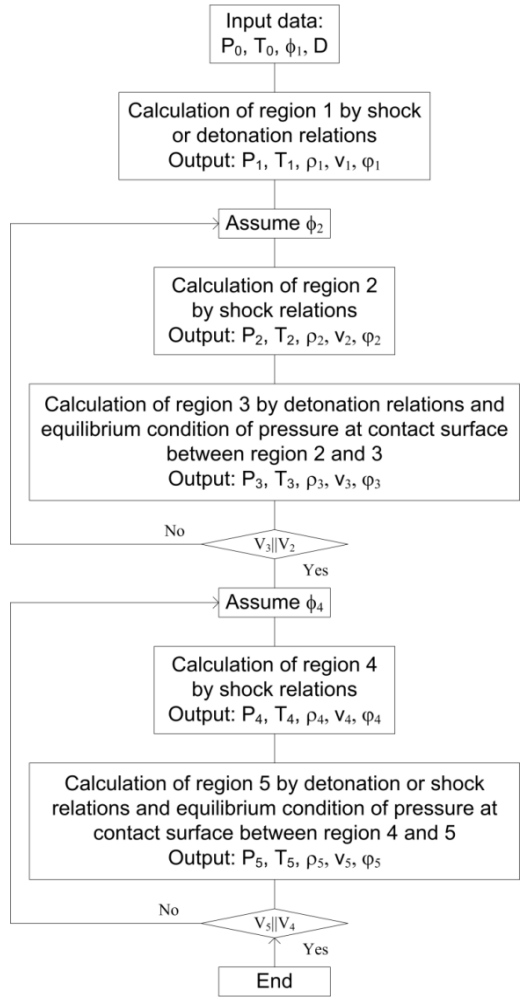

Fig. 10

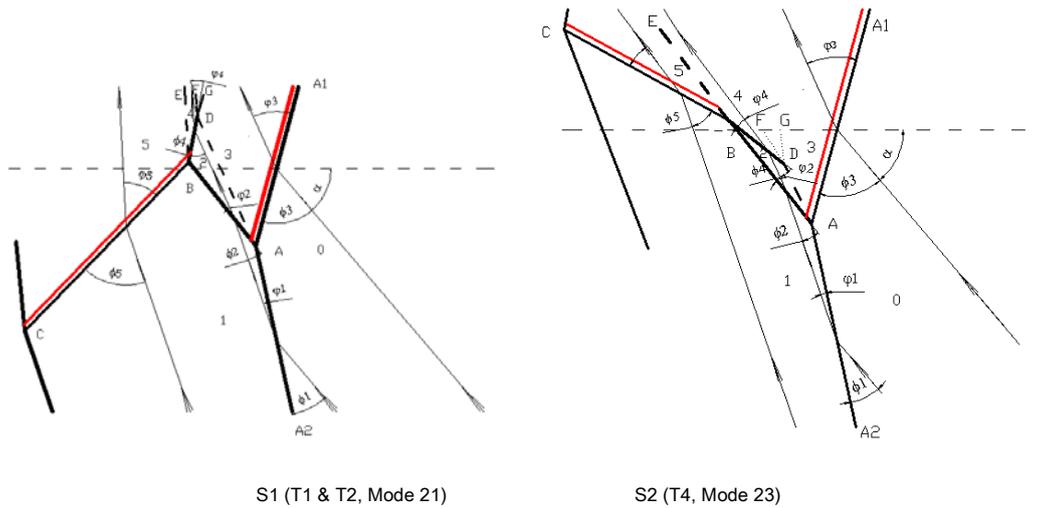

S1 (T1 & T2, Mode 21)　　　　　　S2 (T4, Mode 23)

Fig. 11

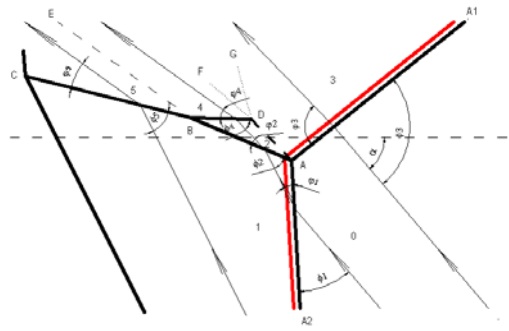

S3 (T3, Mode43)

Fig. 12

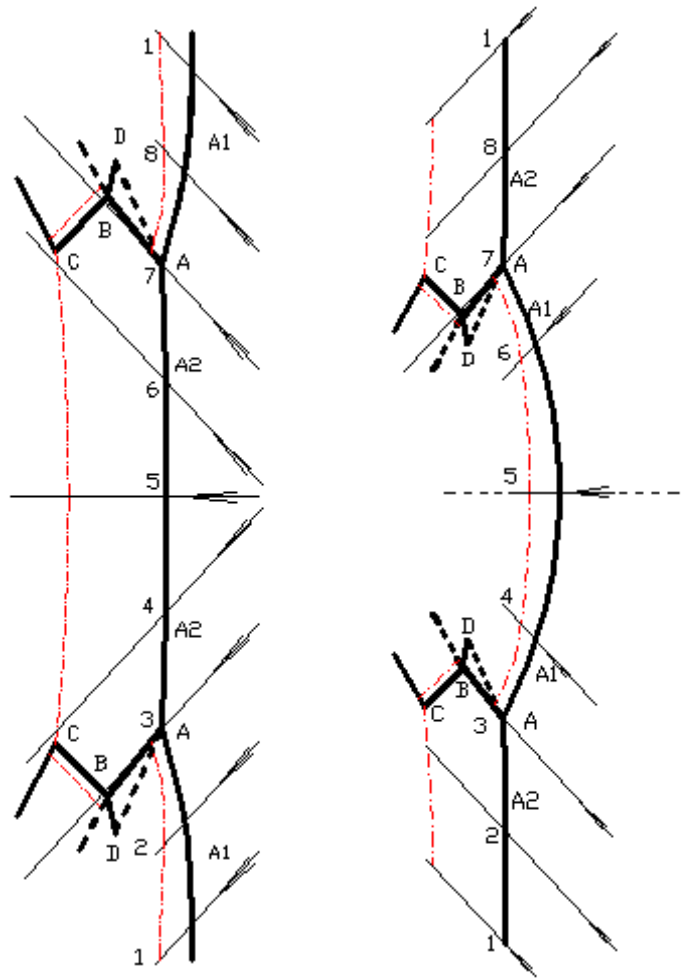

Fig. 13

Table 1

| Structure | Shocks | Region behind shocks | Pressure/ Mpa | | Temperature / °K | Velocity / m/s | Inlet flow angle with shock/ ° | Outlet flow angle with shock/ ° |
|---|---|---|---|---|---|---|---|---|
| | | | Calculation | experiment | | | | |
| S1 | AA2 | 1 | 2.32 | 2.80 | 919.92 | 1998.13 | 32.5 | 6.9 |
| | AB | 2 | 3.23 | / | 988.50 | 1949.88 | 19.93 | 15.56 |
| | AA1 | 3 | 3.23 | 3.0 | 3778.46 | 1581.95 | 63.05 | 47.64 |
| | BD | 4 | 12.25 | / | 1420.39 | 1651.62 | 35.16 | 15.16 |
| | BC | 5 | 12.25 | 12.15 | 2680.3 | 1357.80 | 62.09 | 46.47 |
| S2 | AA2 | 1 | 2.32 | 2.9 | 919.92 | 1998.13 | 32.5 | 6.9 |
| | AB | 2 | 3.23 | / | 988.50 | 1949.88 | 19.93 | 15.56 |
| | AA1 | 3 | 3.23 | / | 3778.46 | 1581.95 | 63.05 | 47.64 |
| | BD | 4 | 7.20 | / | 1200.21 | 1809.83 | 26.53 | 15.44 |
| | BC | 5 | 7.20 | 7.1 | 1575.48 | 1652.09 | 42.65 | 27.19 |
| S3 | AA2 | 1 | 1.71 | / | 1964.72 | 1989.15 | 40 | 25.07 |
| | AB | 2 | 4.12 | 3.8 | 2541.76 | 1656.90 | 40.25 | 23.61 |
| | AA1 | 3 | 4.12 | 4.0 | 4748.42 | 1313.36 | 87.84 | 86.13 |
| | BD | 4 | 5.73 | / | 2647.79 | 2647.79 | 42.02 | 35.00 |
| | BC | 5 | 5.73 | 6.0 | 2719.99 | 2719.99 | 49.13 | 25.47 |